\documentclass[useAMS]{mn2e}
\usepackage{graphicx}
\title[GBS radio sources]{Radio sources in the Chandra Galactic Bulge survey\footnote{Based on observations made with ESO Telescopes at the La Silla Paranal Observatory under programme ID 085.D-0441 and 087.D-0596}}
\author[Maccarone et al.]{Thomas J. Maccarone${^1}$,  Manuel A.P. Torres $^{2,3}$, 
\newauthor Christopher T. Britt$^4$, Sandra Greiss $^5$, Robert I. Hynes$^4$, Peter G. Jonker$^{2,3,6}$,   
\newauthor Danny Steeghs$^{5,3}$, Rudy Wijnands$^7$, Gijs Nelemans$^{6,8}$\\
$^1$Faculty of Physical and Applied Sciences, University of Southampton, Hampshire, SO17 1BJ\\
$^2$SRON, Netherlands Institute for Space Research, Sorbonnelaan 2, 3584 CA, Utrecht, The Netherlands\\
$^3$Harvard-Smithsonian Center for Astrophysics, 60 Garden Street, Cambridge, MA, 02138, USA\\
$^4$Department of Physics and Astronomy, Lousiana State University, Baton Rouge, LA, 70803-4001, USA\\
$^5$Astronomy and Astrophysics, Department of Physics, University of Warwick, Coventry CV4 7AL\\
$^6$Department of Astrophysics, IMAPP, Radboud University Nijmegen, Heyendaalseweg 135, 6525 AJ, Nijmegen, The Netherlands\\
$^7$Astronomical Institute ``Anton Pannekoek'', University of Amsterdam,  Postbus 94249,1090 GE, Amsterdam, the Netherlands\\
$^8$Institute for Astronomy, K.U. Leuven, Celestijnenlaan 200D, 3001 Leuven, 
Belgium\\}

\begin{document}
\def\ltsim{\mathrel{\rlap{\lower 3pt\hbox{$\sim$}}
        \raise 2.0pt\hbox{$<$}}}
\def\gtsim{\mathrel{\rlap{\lower 3pt\hbox{$\sim$}}
        \raise 2.0pt\hbox{$>$}}}

\date{}

\pagerange{\pageref{firstpage}--\pageref{lastpage}} \pubyear{}

\maketitle

\label{firstpage}

\begin{abstract}

We discuss radio sources in the Chandra Galactic Bulge survey region.
By cross-matching the X-ray sources in this field with the NVSS
archival data, we find 12 candidate matches.  We present a
classification scheme for radio/X-ray matches in surveys taken in or
near the Galactic Plane, taking into account other multi-wavelength
data.  We show that none of the matches found here is likely to be due
to coronal activity from normal stars because the radio to X-ray flux
ratios are systematically too high.  We show that one of the sources
could be a radio pulsar, and that one could be a planetary nebula, but
that the bulk of the sources are likely to be background active
galactic nuclei (AGN), with many confirmed through a variety of
approaches.  Several of the AGN are bright enough in the near infrared
(and presumably in the optical) to use as probes of the interstellar
medium in the inner Galaxy.
\end{abstract}


\section{Introduction}

The {\it Chandra} Galactic Bulge Survey (GBS) is a 12 square degree,
shallow (2 ksec depth) survey of the inner Galaxy (Jonker et al. 2011
-- hereinafter J11).  The depth of the survey is chosen so that a
large number of X-ray binaries can be detected, without the X-ray
binaries being overwhelmed in numbers by cataclysmic variables and
coronally active stars.  The survey consists of two $6\times1$ square
degree strips, centered at zero degrees Galactic longitude and
$\pm$1.5 degrees Galactic latitude -- the densest star fields that are
not heavily affected by extinction.  The goals of the survey are to
understand the Galactic field's faint X-ray source population and to
identify interesting individual objects for further follow-up.  Some
examples of the latter are eclipsing low mass X-ray binaries for which
precise masses can be measured.  At the present time, results have
been reported for the first 9 square degrees of the GBS (J11).

The X-ray sources in this survey should form a heterogeneous
population.  A simple population synthesis calculation presented in
J11 for the full 12 square degrees predicted that the
survey would detect about 1600 sources, of which about 700 would be
non-compact stars (mostly RS~CVn and W~UMa stars), about 600 would be
cataclysmic variables, and about 300 would be X-ray binaries.  Smaller
numbers of sources are expected in a wide variety of other classes
(e.g. background active galactic nuclei and millisecond pulsars).  The
first 9 square degrees of the survey turned up 1234 X-ray sources, so
the total number density of sources from the population synthesis
modeling is in good agreement with the real data.

Significant amounts of multi-wavelength follow-up are needed to
determine the actual fractions of sources in each category, and to
identify the most interesting individual sources.  Substantial
progress can be made on this front by using public catalogs, and in
this paper, we present the results of matching the GBS sources against
the National Radio Astronomy Observatory (NRAO) Very Large Array (VLA)
Sky Survey (NVSS; Condon et al. 1998) catalog, and what these sources
are likely to be.  A wide variety of Galactic and extragalactic
classes of objects can emit in both the radio and X-ray bandpasses,
and in many cases, just the ratio of radio flux to X-ray flux can
dramatically limit the range of possible classes of objects, and
additional constraints can come from adding photometric data from
public catalogs.  We also consider correlations against other radio
surveys, mainly to add spectral information or data from higher
spatial resolution imaging, but we use the NVSS matches as the
baseline for our work on radio correlations with the X-ray sources
because it is the deepest survey covering the entire GBS field.  

\section{The data}
We take the GBS catalog from Jonker et al. (2011).  Throughout this
paper, we will refer to sources as CX~$N$, where $N$ is the number of
the entry in the GBS catalog, rather than using a more cumbersome name
based on the source's position on the sky.  We also note that the GBS
catalog is sorted by numbers of source counts, so that the low number
sources in the GBS catalog are the brightest ones, which may lead to
the source names having a bit more value in conveying some intuition
about their properties to the readers.  We convert the X-ray count
rates to fluxes from 0.5-8.0 keV using the same relation as in J11,
$8\times10^{-15}$ ergs/sec/cm$^2$/photon, which comes from assuming a
power law spectrum with $\Gamma=2$ and foreground absorption with
$N_H$=$10^{22}$ cm$^{-2}$.  The counts-to-energy conversion is likely
to produce no more than a factor of a few systematic uncertainty in
the X-ray luminosity for the different classes of sources that could
be emitting both in radio and X-rays, on the same order as the
statistical uncertainties for the faintest sources in the survey.  We
also ignore the vignetting corrections for the sources near the edges
of the field of view, but these corrections are also likely to be no
more than a $\sim10\%$ for off-axis angles less than 10' (see
Beckerman et al. 2003).

The NVSS covers the entire sky north of a declination of $-40$ degrees
at 1.4 GHz, and reaches a sensitivity level of about 2.5 mJy.  The
survey was obtained in D configuration with the VLA, and hence the
angular resolution of the survey is about 45''.  There are 629 radio
sources in the NVSS in the full GBS survey region -- so the NVSS
source density is a bit less than half as high as the X-ray source
density in the region.  The positional accuracy of the bright NVSS
sources is often better than 1'', but for the faintest sources in the
catalog, it can be significantly worse.  After inspection of a source
list with a large matching radius, we decided to work with sources
matching within 5'', because the number of matches started to fall off
beyond that separation.  There are 12 matches found, and their
properties are listed in Table 1.  All but one of the matches obtained
here (source CX~390) are within either 1.2'' or within the $1\sigma$
positional uncertainty of the NVSS.

The absolute astrometric calibration of the Chandra data is accurate
to about 0.6'', based on matches between well-positioned {\it Chandra}
sources and Tycho catalog sources, so we do not consider it
problematic to include sources with greater than $1\sigma$ positional
offsets, given that the offsets are in all cases within 0.6'' of
$1\sigma$ (and that with 12 real matches, it should be expected that a
few of them will be at greater than 1$\sigma$ in any event).  To
estimate the number of chance superpositions, we shifted the positions
of the X-ray sources by 12'' and then by 24'' both to the north and
south, and attempted matching against the NVSS catalog.  Doing so
resulted in 3 matches in the four runs, so we estimate that there is
likely to be approximately 1 false match in our list of likely
associations.  The sources matching within 1.2'' are highly likely to
be real matches, and probably, at most, one or two of the other
matches is a chance superposition. 

We have also searched against other radio catalogs made in this survey
region.  Where matches have been found, they are discussed within the
descriptions of the individual sources.  No matches were found against
the survey made by Langston et al. (2000) with the NASA Green Bank
Earth Station, which reached a flux density of 0.9 Jy at 8.35 GHz and
2.5 Jy at 14.35 GHz.

\begin{table*}
\begin{tabular}{|r|l|r|r|r|r|r|r|r|r|r|l|r|r|r|r|}
\hline \multicolumn{1}{|c|}{GBS} & \multicolumn{1}{c|}{CXOGBS}
  & \multicolumn{1}{c|}{RA} & \multicolumn{1}{c|}{DEC} &
  \multicolumn{1}{c|}{$N_X$} & \multicolumn{1}{c|}{$\sigma_{GBS}$} &
  \multicolumn{1}{c|}{$\sigma_{NVSS}$} & \multicolumn{1}{c|}{NVSS} &
  \multicolumn{1}{c|}{S1.4} & \multicolumn{1}{c|}{e\_S1.4} &
  \multicolumn{1}{c|}{Separation} &
  \multicolumn{1}{c|}{$K_s$}\\ \hline 
2 & J173728.3-290802 &   264.36831 &-29.13389 & 2191 &0.1 & 0.8 & 173728-290801 & 47.4 & 1.5   & 0.1 & 10.94\\ 
40 & J174404.3-260925 & 266.01795 & -26.1571 & 35 & 0.1
  & 2.4 & 174404-260924 & 7.3 & 0.5 & 0.7 & 13.56\\ 
49 & J173146.8-300309 &
  262.94521 & -30.05255 & 30 & 0.2 & 0.8 & 173146-300309 & 90.5 & 2.8
  & 0.5 & 13.45\\ 
52 & J174423.5-311636 & 266.09819 & -31.27687 & 29 & 0.5 &
  0.8 & 174423-311636 & 424.6 & 14.9 & 0.2 & ND\\ 
233 & J174206.1-264117 &
  265.52569 & -26.68812 & 10 & 1.4 & 7.8 & 174206-264119 & 3.5 & 0.5 &
  5.0 & 15.68?\\ 
293 & J174000.6-274816 & 265.00267 & -27.8046 & 9 & 0.2 & 0.7
  & 174000-274816 & 169.5 & 6.0 & 0.4 & 13.33\\ 
390 & J173607.5-294858 &
  264.03134 & -29.81623 & 7 & 0.4 & 1.0 & 173607-294855 & 29.0 & 1.3 &
  3.1 & 15.92\\ 
488 & J173605.3-283232 & 264.02231 & -28.54229 & 6 & 0.5 & 0.8
  & 173605-283232 & 125.8 & 4.5 & 1.1 & 14.94\\ 
494 & J173458.8-301328 &
  263.74511 & -30.22469 & 6 & 0.8 & 1.0 & 173458-301329 & 28.6 & 1.0 &
  0.9 & 12.51?\\ 
578 & J174442.3-311633 & 266.17636 & -31.2761 & 5 & 0.3 & 14.1
  & 174442-311637 & 2.8 & 0.6 & 3.1 & 14.19?\\ 
937 & J175359.2-281720 & 268.49683 & -28.28917 & 3 & 0.2 & 0.8 & 175359-281722 & 51.9 & 1.6 &  1.0 & 13.32\\ 
1234 & J173531.4-295145 & 263.88121 & -29.86272 & 3 & 0.5 &
  5.0 & 173531-295147 & 4.7 & 0.5 & 3.3 & 15.84\\ \hline\end{tabular}
\caption{The matches between the Chandra GBS sources and the NVSS
catalog.  The columns are: (1) GBS source number (2) IAU catalog name
for the GBS sources (3) right ascension in decimal degrees for the GBS
sources (4) declination in decimal degrees for the GBS sources (5)
number of X-ray counts detected in the GBS data (6) positional
uncertainty in arcseconds in the GBS (7) positional uncertainty in
arcseconds in NVSS (8) NVSS catalog entry (9) flux density of the NVSS
sources in mJy (10) uncertainties on the NVSS flux densities
in mJy (11) separations between the NVSS source and Chandra
source positions in arcseconds (12) the $K_s$ magnitude of the source, mostly from VVV (see text for cases where the values come from 2MASS), with question marks next to the numbers for sources with separations more than 1.5'').  We note that source CX~2 is in two different pointings of the GBS, and is also piled-up, so its ratio of total number of counts to X-ray flux is different from that of all the other sources (and is not well measured because of the pile up.)}
\end{table*}

\begin{table*}
\begin{tabular}{lrrrrl}
\hline
Source class&log ($L_R/L_X$)& $L_X$ & $M_K$ & Radio spectrum & Comments\\
\hline
Active star & -15.5 & $10^{26}-10^{32}$ & wide range & $\approx$ flat & saturates at $L_x=10^{-3} L_{bol}$\\
&&&&& highest $L_X$ for RS CVn only\\
Ultracool active dwarf & -15.5 to -11.5 & $\ltsim10^{26}$ & $\approx10$& $\approx$ flat& \\
Colliding wind binary&$\sim-14$&$\sim10^{32-34}$&-5 to -7&$\approx$flat&\\
Single massive star&$\sim-14$&$10^{31-32}$&-5 to -7&+0.6&\\
Pre-main sequence star&-15 to -13& $\ltsim 10^{31}$&$\ltsim 0$& usually flat& based on highly incomplete sample\\
Class I YSO& -15 to -13& $\ltsim 10^{31}$&$\ltsim 0$& usually +0.6& based on highly incomplete sample\\
&&&&& some outliers exist in spectral index\\
Class 0 YSO&&&&&undet. in X-rays\\
X-ray binary&-15 to -13*&$10^{30-39}$&wide range& flat at low $L_X$& *radio can be brighter for some HMXBs\\
&&&&&much fainter for neutron star systems,\\
&&&&&espec. faint for slow X-ray pulsars\\
Pulsar&-16.7 to -6.7 & $10^{25}$to $10^{37}$&faint&Steeper than -1&X-rays from Possenti relation,\\
&&&&&so calibrated only above $L_X=10^{29};$\\
&&&&&can have bright $K$ if massive companion\\
Symbiotic star&$\sim-12$ to -11&$10^{30-33}$&-3.4 to -8&inverted&\\
SNR&&&&&extended for any reasonable age;\\
&&&&&other properties widely variable;\\
&&&&&known Gal. SNRs $\gtsim$ 1~Jy\\
PNe&$>-11.5$&$\sim10^{30}$& 0 to -3 &usu. flat& Opt., IR are line dominated\\
PWN&-10&$>10^{34}$&faint&flat& oft. extended\\
AGN&-9.7 to -14.9&$\ltsim10^{46}$&wide range&usu. -0.7&rough IR, $L_X$ correlation, oft. ext. lobes\\
Starburst&-10.5&$10^{40-42}$&wide range&-0.7&IR $\sim$ w/$L_X,L_R$\\
Cluster of galaxies& -13 to -14& $10^{44.5}-10^{45.5}$& -25.5 & steeper than -1& mag for brightest indiv. gals.\\
\hline
\end{tabular}

\caption{A scheme to classify radio/X-ray matches in Galactic survey
fields, which also includes possible extragalactic classes of objects,
since these will often represent a large fraction of the total source
counts for radio/X-ray matches.  The columns are source class, base 10
logarithm of the ratio of radio flux density to X-ray flux in units of
Hz$^{-1}$, X-ray luminosity in ergs/sec, absolute $K$ band magnitude,
a description of the radio spectrum, with spectral indices given as $\alpha$, where $F_\nu \propto \nu^{-\alpha}$, and a column for other comments about what information can be used to identify sources in these classes.  The abbreviations in the table are as follows: YSO = young stellar object; PNe = planetary nebulae;  SNR = supernova remnant; PWN = pulsar wind nebula; AGN = active galactic nucleus; HMXB = high mass X-ray binary}
\end{table*}

\subsection{Optical spectroscopy}
We have obtained spectra of the optical counterpart to CX~40 using
the VIsible Multi-Object Spectrograph (VIMOS), the 4-channel imager
and spectrograph mounted on the Nasmyth focus on the 8.6-m ESO Unit 3
Very Large Telescope at Paranal, Chile (Le F\`evre et al., 2003).  For
the observations we used VIMOS in MOS mode and selected the MR grism
that cover the wavelength range $4900-10150$ \AA~ and yields a 2.5
\AA/pixel dispersion.  A 1.0'' arcsec slit width laser-cut at the
source position (R.A = 266.018000, Dec = -26.156984) provided a
spectral resolution of 750.

The observations for CX~40 were performed as a 1 hr service mode
observing block (OB) on the night of 29 April 2011 (JD 2455680; OB ID
509202).  The OB contained two spectroscopic integrations of 875~s on
source, three flat-field exposures and arc lamp exposure for
wavelength calibration .  The data reduction was performed with the
ESO-VIMOS pipeline.

\subsection{Optical and infrared photometry}
We acquired 8 nights of photometry, from 12-18 July 2010, with
the Blanco 4.0 meter telescope at the Cerro Tololo Inter-American
Observatory (CTIO). Using the Mosaic-II instrument, we observed the 9
square degree area containing the X-ray sources identified by the GBS
(Jonker et al. 2011). Multiple SDSS $r'$ band exposures with an
integration time of 120 s of 45 overlapping fields were taken to cover
the area. Typical seeing for the run was around 1''. The order in
which each field was observed was randomized to minimize aliasing
based on the time it took to cycle through fields. The data were
reduced via the NOAO Mosaic Pipeline, which also added a world
coordinate system to the images.  The magnitudes produced are
calibrated to USNO-B1 stars in the field by the NOAO Mosaic
Pipeline. These stars are not carefully calibrated standards, and the
magnitude calibration based on them carries an uncertainty of 0.5
magnitudes as quoted by the NOAO Data Handbook.

Photometry was done with Alard's ISIS version 2.2, detailed in Alard
\& Lupton (1998) and Alard (2000). ISIS works by using a reference
image which it then convolves with a kernel in an effort to match a
subsequent image of the same field. The subsequent image is then
subtracted from the convolved reference image. Since most stars are
not variables, and are affected by atmospheric conditions similarly as
nearby stars in the field, the convolution can match non-variable
field stars from image to image.  Therefore, any residual flux after
subtraction is due to an inherent change in the brightness of a
star. In order to save computation time, postage stamp sized cutouts
of the full Mosaic images were taken around each object for
processing. Any variables inside the X-ray position confidence region
are considered candidate optical counterparts, though we only expect
to see $\sim$20-25 chance alignments of variable optical stars and X-ray positions, given typical crowding in the fields and the abundance of variable field stars.

The result of the ISIS routine is a series of flux changes from the
reference image, which are then converted to magnitudes by using IRAF
to determine the number of counts from the star of interest in the
reference image.  Despite being in the unusual units of a change in
flux, the photometry is differential because the convolution is done
such that all non-variable field stars are matched between images,
which has the same effect as normalizing counts to flux from
comparison stars. Essentially, every star in the field becomes a
comparison star.

Infrared photometric data are taken from existing catalogs.  For
relatively faint sources, we use the Vista Variables in the Via Lactea
(VVV) survey data (Minniti et al. 2010; Saito et al. 2010; Saito et
al. 2012), with flux from with the standard 2'' apertures used.  The
uncertainties on the photometry are generally greater than the
differences between the flxues from the 1'' and the 2'' apertures.
For brighter sources, in which VVV can be saturated, we also look at
the data from the 2MASS survey (Cutri et al. 1998).  Details for which
catalog has been used are given in the notes on the individual
sources.

\subsection{Radio to X-ray flux ratios}

It is useful to look at the radio to X-ray flux ratios for these
objects in order to be able to make a quick first cut on the classes
of sources possible.  Radio power is typically presented as a
luminosity density (i.e. in ergs/sec/Hz), with most surveys for which
standard relations are calibrated taking place at 5 GHz.  Instead, we
provide ratios between the X-ray luminosity and the luminosity density
at 1.4 GHz, but note that the differences between the two are likely
to be no more than half an order of magnitude over the range of
spectral shapes typically observed in the radio -- since many of the
X-ray sources have fewer than 10 counts, the $1\sigma$ statistical
errors on the observed ratio of X-ray flux to radio flux density will
typically be larger than the uncertainties introduced due to the
differences in flux density at 1.4 and 5 GHz.  X-ray power is
typically given as an actual flux within a broad bandpass rather than
as a flux density.  We plot the raw data, the X-ray fluxes versus
radio flux densities, in Figure 1, rather than plotting the flux
ratios.  The values of $L_R/L_X$ are typically in the $10^{-10}$ to
$10^{-12.5}$ Hz$^{-1}$ range for these matches.
 
The data from which we find standard relations for different source
classes between X-ray and radio luminosities come from a variety of
sources using a variety of satellites.  As a result, some measurements
will be well matched to the 0.5-8.0 keV bandpass we use, while others,
especially those data taken with ROSAT, will be in somewhat different
bandpasses.  Most of the classes of sources under consideration have
spectral shapes not too different from the absorbed power law with
$\Gamma=2.0$ that we use to estimate luminosities for our {\it
Chandra} sources.  With such a spectral shape, the corrections between
e.g. the ROSAT bandpass of 0.1-2.4 keV and the Chandra bandpass of
0.5-8.0 keV are about 15\%.  The thermal spectra of coronally active
stars also have bandpass corrections of less than about 10\% for
different satellites (G\"udel \& Benz 1993).  As a result, we simply
quote and use the X-ray luminosities in the bands for which the
relations for the individual source classes were derived, but note
that some caution should be used in excluding source classes simply
because of a ratio of radio flux density to X-ray flux that is just
outside the quoted boundaries for a source class.

\begin{figure}
\includegraphics[width=8 cm]{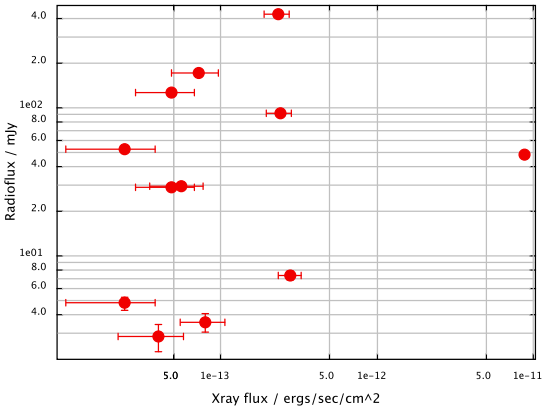}
\caption{The radio flux densities plotted versus the X-ray fluxes of the GBS-NVSS matches, in a log-log plot.  A 10 mJy radio source at $10^{-13}$ ergs/sec/cm$^2$ would have a $L_R/L_X$ of $10^{-12}$ Hz$^{-1}$.  The flux of the brightest X-ray source is likely to be an underestimate, because its count rate is affected by pile-up.}
\end{figure}

\section{Possible sources of radio and X-ray emission}
A variety of classes of objects can be found to emit in both radio and
X-rays, especially when the X-ray fluxes observed are rather faint.
We will present a summary of the different classes of sources which
are known to emit in both bands, and the typical flux ratios between
radio and X-rays, as well as the typical characteristic luminosities.
We also consider how data from other wavelengths (especially infrared
and optical) can help characterize these sources.  We present a
discussion even of source classes which can be ruled out in this
survey in part to create a useful resource for future studies which
might have different relative depths among the X-ray, radio, and
optical/infrared catalogs.  Because we are focused on Galactic objects
in general and the inner Galactic Bulge for the sources in this paper,
we focus primarily on the infrared, rather than optical, fluxes, since
the infrared measurements will suffer much less from extinction than
the optical fluxes.

\subsection{Non-compact stars}
\subsubsection{Coronally active stars}
Relatively well established relations between X-ray and radio power
exist for a few classes of sources.  For coronally active stars, in
which the X-ray and radio emission come from magnetically powered
coronae on rapidly rotating late-type stars, a linear (or perhaps
nearly linear) relation between X-ray and radio power has been
well-established for quite some time (G\"udel \& Benz 1993).
Typically, it is found that ${\rm log} (L_R/L_X) = -15.5\pm0.5$ Hz$^{-1}$, with the radio luminosity densities usually taken at 6 cm and the X-ray luminosities measured predominantly with {\it ROSAT} (0.1-2.4 keV) or
{\it Einstein} (0.4-4 keV).  The radio spectra are usually flat enough
that the choice of frequency is not very important.  The relation is
the same for coronally active stars which are young single stars, and
for older active stars in binaries.  The W~UMa stars, a class of
mass-transferring binaries, are systematically underluminous in both
radio and X-rays relative to coronally active binaries with similar
orbital periods (Crudacce \& Dupree 1984; Rucinski 1995).  Algol
systems are also thought to have their X-ray and radio emission
dominated by coronal activity (e.g. Favata et al. 2000).

\subsubsection{Coronally active ultracool dwarfs}
Ultracool dwarfs (stars later than M6), appear to be overluminous in
the radio relative to the X-rays by several orders of magnitude
(e.g. Berger et al. 2010).  It is thus possible for ultracool dwarfs
to have X-ray/radio ratios in the range observed here.  On the other
hand, the X-ray luminosities of these objects are quite small -- less
than about $10^{26}$ ergs/sec, and more typically less than $10^{25}$
ergs/sec, meaning that the GBS will be sensitive to them only out to
about 100 pc for the very brightest objects, and in general only
within about 30 pc.  These stars typically have $K$-band absolute
magnitudes of about 10, and $J-K$ colours of about 1.0-1.5, making
them objects which should stand out well in the infrared surveys
already existent in the GBS region (i.e. 2MASS for the relatively
nearby objects -- Cutri et al. 2003; and the Vista Variables in the
Via Lactea -- VVV -- survey for more distant objects -- Minniti et
al. 2010).  Many will additionally have large proper motions (of order
0.2''/year).

\subsubsection{Massive stars}
A variety of classes of massive stars can provide strong X-ray
emission, generally with accompanying radio emission.  Clark et
al. (2008) provide a nice summary of the different classes of X-ray
stars expected in a young massive star cluster, and much of our
discussion is drawn from that work.

In massive stars, the X-rays come from shock-heated gas in the
objects' stellar winds. These shocks can be caused either by
self-shocking of the wind of a single star, or by collisions of the
winds of two massive stars in a binary.  Single massive stars
typically show ratios of $L_X$ to bolometric luminosity of $\approx
10^{-7}$ (Long \& White 1980), with the relation derived using {\it
Einstein} data from 0.15-4.5 keV, so typically $L_X$ will be in the
$10^{31}$-$10^{32}$ ergs/sec range for OB stars.  Colliding wind
binaries are typically more luminous in the X-rays, due to the shocks
from the wind-wind collisions (Pollock 1987), often reaching $10^{34}$
ergs/sec in an 0.8-4.0 keV band with {\it Einstein}.

Radio emission from massive stars has been reviewed by Seaquist
(1993).  The radio emission from single massive stars is thermal, from
the photoionization of the stellar wind by the star, and typically has
$L_R$ of $10^{17}-10^{19.5}$ ergs/sec/Hz at 6 cm, with the inverted
spectrum (typically $f_\nu \propto \nu^{0.6}$) yielding slightly lower
luminosities at 1.4 GHz.  Colliding wind binaries are seen to have
flatter, non-thermal spectra, with radio luminosities typically of
$10^{18.5}-10^{20}$ ergs/sec/Hz (e.g. Bieging et al. 1989).  Both
single OB stars and colliding wind binaries are thus within the range
of $L_R/L_X$ observed in our sample.  

Massive stars will be very bright optical and infrared objects.  One
can take, e.g. the infrared fluxes from 2MASS (Cutri et al. 2003) for
the X-ray detected Wolf-Rayet stars from Westerlund 1 (Clark et
al. 2008), most of which are also radio emitters (Dougherty et
al. 2010).  The apparent magnitudes of these objects range from
$K=7.2$ to $K=9.2$.  Given the typical extinction value $A_K=1.3$ in
Westerlund 1 (Clark et al. 2008; Cardelli et al. 1989), which is a bit
higher than the extinction to the GBS fields, and the distance of
Westerlund 1 of about 5 kpc (Clark et al. 2008), a bit closer than the
Galactic Centre distance, we expect that Wolf-Rayet stars in the GBS
would be no fainter than about $K=10$.  OB stars with strong winds
should be in a similar magnitude range.

\subsubsection{Pre-main sequence stars and young stellar objects}
Both Young Stellar Objects (YSOs) and Pre-Main Sequence (PMS) stars
are known to produce radio emission. Classical (Class~II) and
Weak-line (Class~III) T~Tauri stars can also appear as relatively
bright sources in the X-ray band and detectable sources in the radio.
A small sample of young stellar objects and pre-main sequence stars
has been observed simultaneously in the radio and X-rays by Forbrich
et al. (2011).  They find more than an order of magnitude of scatter
in the ratio of X-ray to radio luminosity for these objects, and
significant numbers of non-detections in both bands.  The typical
sources fall about one order of magnitude brighter in radio for a
given X-ray luminosity than the standard G\"udel-Benz relation for
coronally active stars, so that the $L_R/L_X$ values range from about
$10^{-13.5}$ to $10^{-15.5}$.  The observed sources are mostly in the
$10^{29}$-$10^{31}$ ergs/sec range in X-rays.

Additionally, the high ratio of infrared to radio/X-ray emission for
pre-main sequence stars makes them unlikely to amongst the sources
reported in this paper. Only the nearest PMS stars will have radio
flux densities higher than the NVSS completeness limit; for example,
T~Tau itself has a radio flux density of only a few mJy (Scaife
2011a), and a $K_s$ magnitude of 5.3 (Cutri et~al. 2003). This places
it towards the lower end of the sources found in the NVSS, while its
near infrared flux is about 6 magnitudes brighter than any of the
sources we find with radio counterparts.  A deeper radio survey could
be expected to turn up a substantial fraction of the pre-main sequence
stars in the GBS survey.

Earlier phases of protostellar evolution (Class~0/I) show strong X-ray
emission more rarely (Feigelson \& Montmerle 1999). Class~0 objects
are deeply embedded in dense dust envelopes and have SEDs which peak
in the sub-mm/FIR. Such objects have never been unambiguously detected
in X-rays, although one strong candidate has been found in which the
X-ray emission may result from shocking due to an abnormally fast jet
being launched into the system (Hamaguchi et~al. 2005). Class~I
objects, which have accreted more than half of their dust envelope
onto the central source, are often seen to emit in X-rays ($L_{\rm
X}\leq 10^{31}$\,erg\,s$^{-1}$). Class~I objects tend to have large
ratios of near-infrared (and especially mid- and far-infrared) flux to
X-ray flux, and generally have low radio luminosities, with $L_{\rm
R}<10^{18}$ ergs/sec/Hz at 15 GHz, and their radio spectra are
generally rising, so that the luminosity densities will be even
smaller in the 1.4 GHz NVSS band (Scaife et~al. 2011b;c).  The values
of $L_R/L_X$ for the detected class I young stellar objects span the
same range as for the T Tauri stars, but the class I objects are more
likely to be non-detections in the Forbrich et al. (2011) sample than
are the T Tauri stars.

\subsection{Compact stars}
\subsubsection{X-ray binaries}
Black hole X-ray binaries represent another possible source class of
interest for our work.  Gallo et al. (2003) find that
$\left(\frac{L_R}{10^{18} {\rm ergs/sec/Hz}}\right) = 120
\left(\frac{L_X}{10^{36} {\rm ergs/sec}}\right)^{0.7}$, with the X-ray
luminosities taken from 2-10 keV and the radio luminosity densities
taken at 5 GHz.  For a quiescent black hole X-ray binary at
$L_X=10^{32}$ ergs/sec, the value of $L_R/L_X$ will then be about
$10^{-14.7}$.  Therefore, an X-ray binary would have to be
overluminous in radio relative to the Gallo et al. (2003) relation by
a substantial factor in order to be one of the GBS sources, since the
GBS sources are all a factor of 100 higher in $L_R/L_X$ than the
typical value.

A few binaries with neutron stars or black holes do show radio
emission far in excess of the Gallo et al. (2003) relation -- SS~433,
Cygnus~X-3, and the binaries which also emit high energy gamma-rays
(see e.g. Fender 2006 for a review).  As these systems are all
high-mass X-ray binaries with strong stellar winds, it is likely that
these sources have their jets interacting with stellar winds (or in
the case of the gamma-ray sources, may be powered by the interaction
of a pulsar wind with a stellar wind -- Maraschi \& Treves 1981).
However, these sources are all still very luminous in both radio and
X-rays ($L_X\sim10^{35-36}$ ergs/sec), and hence are not likely to be
relevant to the objects reported in this paper.  The upcoming generation of radio telescopes will be needed to determine whether there exists a population of low luminosity X-ray binaries with radio fluxes well above the predictions of the Gallo et al. (2003) correlation.

It has, on the other hand, become clear that there exists a population
of radio sub-luminous black hole X-ray binaries (e.g. Coriat et
al. 2011), but these sub-luminous sources tend to exist only at
relatively high fractions of the Eddington luminosity, where low mass
X-ray binaries would be very bright X-ray sources in the GBS, even if
they were located far past the distance to the Galactic Center, and
hence would have already stood out in our survey which consists
predominantly of much fainter sources.  As these sources become
fainter, they seem to return to the standard radio/X-ray correlation
(Coriat et al. 2011; Maccarone 2011; Jonker et al 2012; Ratti et al
2012).

At the faint end of the flux distribution, where the X-ray
luminosities alone would not distinguish black hole X-ray binaries
from normal stellar objects, Gallo et al. (2006) found A0620-00 to be
a factor of about 2 brighter in the radio than the extrapolation of
the Gallo et al. (2003) relation, while Calvelo et al. (2010) found
upper limits on two other quiescent black hole X-ray binaries,
GRO~J1655-40 and XTE~J1550-564.  These upper limits are marginally
consistent with the Gallo et al. (2003) relation, and below the level
one would expect if the radio emission were enhanced in quiescence as
was suggested from the original observations of A0620-00.
Miller-Jones et al. (2011) found similarly stringent upper limits on
GRO~J1655-40, GRO~J0422+322 and XTE~J1118+480 with the EVLA.  Neutron
star X-ray binaries at low X-ray flux levels generally lie far below
the X-ray/radio relation for black holes (Migliari \& Fender 2006),
while cataclysmic variables lie further still below that relation
(e.g. K\"ording et al. 2008).

\subsubsection{Pulsars}
Rotation-powered pulsars represent another class of objects which can
display both radio and X-ray emission.  The X-ray luminosities of
pulsars are well correlated with their spin-down powers (Verbunt et
al. 1996; Becker \& Tr\"umper 1997; Possenti et al. 2002), with the
first two papers (which use {\it ROSAT} data from 0.1-2.4 keV) finding
that the X-ray luminosity is typically about 0.1\% of the spin-down
luminosity and the latter (which uses a variety of observatories, and
converts luminosities to a 2-10 keV band) finding a similar typical
value, but with a steeper slope, such that the highest spin-down power
pulsars are a bit brighter than 0.1\% of spin-down and the lowest
spin-down power are X-ray fainter than 0.1\% of spin-down.  The radio
luminosities are not well correlated with the spin down powers,
leading to substantial scatter in the ratio of X-ray to radio fluxes
for pulsars.  If, for example, one plots spin-down luminosity versus
radio luminosity for a large sample of pulsars (e.g. Manchester et
al. 2005), one sees approximately six orders of magnitude in scatter.
The parameter space easily covers sources over the full range in flux
densities from 1-100 mJy, and observed (or expected, based on taking
0.1\% of the spin down luminosity) X-ray fluxes in the
$10^{-14}$-$10^{-12}$ ergs/sec/cm$^2$ range for distances in the
$\sim$1-10 kpc range -- the X-ray and radio fluxes alone cannot
typically rule out a pulsar nature for GBS sources.  Pulsars only
rarely have bright optical counterparts, even when they are in
binaries, since most binary pulsars are millisecond pulsars whose
companions are in the very late stages of stellar evolution after
having already donated most of their masses to spin their companions
up (van Kerkwijk et al. 2005).  As a result a radio and X-ray source
with no counterparts at other wavelengths can be a good candidate for
being a pulsar, although the detection of a bright optical or infrared
counterpart does not immediately rule out a pulsar nature for the
X-ray and radio emission, since a small fraction of rotation powered
pulsars do have high mass stars as counterparts.

\subsubsection{Symbiotic stars}
Symbiotic stars represent a final class of compact binary which can
show strong emission at both radio and X-rays.  These are typically
binary systems containing a red giant or asymptotic giant branch star
plus a white dwarf or a neutron star.  Symbiotic stars show strong
thermal radio emission with $L_R\approx 10^{17}-10^{22}$ ergs/sec/Hz
(Seaquist et al. 1993).  X-ray emission from symbiotic stars spans a
wide range in luminosities, from about $4\times10^{29}$ ergs/sec to
about $10^{33}$ ergs/sec for the white dwarf accretors which are not
supersoft sources (M\"urset et al. 1997).  Much higher luminosities
($\gtsim10^{36}$ ergs/sec) are seen for the supersoft sources
(i.e. the sources whose luminosities are expected to be dominated by
steady nuclear burning on the surface of the white dwarf -- Greiner
2000) and for the symbiotic X-ray binaries (i.e. those with neutron
star accretors), which typically have $L_X \gtsim 10^{35}$ ergs/sec
(Corbet et al. 2008).  Symbiotic stars can thus often come within the
range of $L_R/L_X$ observed for our matches.  Bright infrared
counterparts could be taken as additional suggestive evidence for
symbiotic stars -- the absolute $K$-band magnitudes of the symbiotic
stars with Hipparcos parallaxes range from -0.2 to -8.1, with most of
the objects at the bright end of the range (Munari et al. 1997).
These absolute magnitudes correspond to apparent magnitudes in the
range from about 6 to 14 at the distance of the Galactic Centre, with
most symbiotic stars expected to have $K<11.5$.  We note that the
faintest object in the sample, AR~Pav, also has a very poorly measured
parallax from Hipparcos, and that excluding it leaves the faintest
object at $M_K=-3.4$, which corresponds to $K=11.4$ at the Galactic
Center distance.

\subsection{Extended Galactic objects}
\subsubsection{Supernova remnants and pulsar wind nebulae}
Two cases for diffuse Galactic emission sources associated with
compact objects may also produce strong radio and X-ray emission:
supernova remnants and pulsar wind nebulae.  Supernova remnants can be
ruled out as candidates for being the GBS sources on the basis of
their large angular sizes (and younger, more compact supernova
remnants, which have smaller sizes, can be ruled out on the basis of
the relative faintness of these sources -- see e.g. Williams et
al. 1999; Park et al. 2002).  Also, nearly all known Galactic
supernova remnants have radio flux densities of at least 1 Jy around 1
GHz (e.g. Green 2009).  Pulsar wind nebulae can often be found to have
roughly the right ratio of radio to X-ray luminosities, but pulsar
wind nebulae are generally only found around the youngest pulsars,
which are more X-ray luminous than all the sources with radio matches
except CX~2 (e.g. Gaensler \& Slane 2006), and CX~2 is a known active
galactic nucleus (Marti et al. 1998 -- see further discussion in
section 4.1).

\subsubsection{Planetary nebulae}

Planetary nebulae are well known emitters of free-free radio emission
(Condon \& Kaplan 1998), with most of the planetary nebulae from the
Strasbourg catalog (Acker et al. 1992) north of $\delta=-40$ degrees
being detected in the NVSS.  A fraction of planetary nebulae are also
seen as X-ray sources (Guerrrero et al. 2006), but typically these are
found only with dedicated pointings, rather than all-sky X-ray
surveys.  As a result, some rather severe selection effects exist, and
while one can probably make a reasonably fair estimate of the minimum
$L_R/L_X$, it is difficult to estimate the maximum value of
$L_R/L_X$. The most likely origin of the X-ray emission from planetary
nebulae seems to be coronal activity of the binary companions to the
white dwarfs, yielding typical X-ray luminosities of $\sim10^{30}$
ergs/sec (Montez et al. 2010).  The X-ray luminosities are typically
much less than the ultraviolet luminosities, and are therefore are
energetically unimportant for powering the free-free thermal emission
from the nebulae, and hence should be largely uncorrelated with the
radio emission.  From the sample of 6 objects located north of
$\delta=-40$ which have been observed by {\it Chandra} or {\it XMM},
we find that the range of $L_R/L_X$ goes from $3\times10^{-12}$ to
$4.5\times10^{-10}{\rm Hz}^{-1}$, but we note again that the upper end to the range is limited severely by the X-ray sensitivity.  The optical and infrared magnitudes of planetary nebulae will typically be dominated by their line emission, and the $K_s$ magnitudes from 2MASS for Galactic planetary nebulae range from -5 to +5, with the vast majority of objects from -3 to 0 (we derive these by matching the 2MASS
planetary nebula catalog of Ramos Larios \& Phillips 2005 with a
compilation of distances for these objects from Cahn et al. 1992).

\subsection{Extragalactic sources}
\subsubsection{Active galactic nuclei}
The primary extragalactic source for both radio and X-ray emission is
active galactic nuclei, although both clusters of galaxies and
starburst galaxies can be strong emitters at both radio and high
energies as well.  Active galactic nuclei span a large range in the
ratio of X-ray to radio fluxes, whether one focuses on core
luminosity, extended luminosity, or, as is the case with low angular
resolution data at 1.4 GHz, like the NVSS data, mixes the two.

One can take the results of the REX survey (Caccianiga et al. 1999),
to investigate the range of parameter space spanned by active galactic
nuclei in X-ray to radio flux ratios.  The REX project combined the
NVSS sample with that of the ROSAT all-sky survey sample.  The survey
is over a wide field of view, and reaches a flux limit only a factor
of about 10 less sensitive than that of the GBS (albeit with much
worse angular resolution), and so represents a good comparison to the
range we should expect for GBS-NVSS correlations.  Converting the
values of the spectral slopes from radio to optical and from optical
to X-ray from Cacciniga et al. (1999) to the units we use here, we
find that active galactic nuclei should span a range from $10^{-9.7}$
to $10^{-14.9}$ Hz$^{-1}$ in $L_R/L_X$ -- although the real possible
range is probably slightly wider because there will be some AGN
detected only as radio sources or only as X-ray sources.  We note also
that in some cases, the extended structure for AGN may be partially
resolved out by interferometers.  This is not a serious problem for
our purposes in this paper, since we have found the possible range of
$L_R/L_X$ using NVSS data, just as we use NVSS data for the work in
this paper, so the effects of over-resolution should be similar for
our work as for the work of Caccianiga et al. (1999).

Watanabe et al. (2003) have investigated the range of X-ray to near
infrared flux for a large sample of active galactic nuclei with X-ray
fluxes from 2-10 keV above about $2\times10^{-13}$ ergs/sec.  They
find a flux ratio of about 3 for most of the AGN, but also find values
as low as 0.01 and as high as 30.  Therefore, the infrared magnitude
is not, in and of itself, a good indicator of whether a source is
likely to be an active galactic nucleus.

\subsubsection{Starburst galaxies}
Starburst galaxies follow relatively clear relationships between radio
and X-ray fluxes.  In the radio, it has been shown that the star
formation rate is $5.9\times10^{-29} L_{1.4} M_\odot$~yr$^{-1}$, where
$L_{1.4}$ is the 1.4 GHz luminosity density of the galaxy in
ergs/sec/Hz (Yun, Reddy \& Condon 2001).  In the X-rays, it has been
shown that the star formation rate of a galaxy in solar masses per
year is well correlated with its X-ray luminosity, with
$L_{2-10}/6.7\times10^{39} {\rm ergs/sec}$ (Grimm et al. 2003) fitting
the relation, albeit with large scatter and a non-linear best-fit
relation at low luminosities due to the stochastic effects of small
numbers of bright X-ray sources.  The typical $L_R/L_X$ for a
starburst galaxy will thus be about $3\times10^{-12}$.  The starbursts
would have to be relatively compact or at relatively large distances
in order not to cause the X-ray emission either to be resolved, or to
be so extended that the X-ray sources become undetected (as might
happen at the faint end of the X-ray flux distribution).
Additionally, the starbursts are a bit too X-ray bright to match well
with the objects in our sample, although an outlier at low $L_R/L_X$
could fit with the more radio quiet matches in the sample.

Starburst galaxies can be expected to have relatively bright $K$ band
emission, albeit with much scatter.  We can take M~82 as a local
example of a galaxy with a high specific star formation rate, and
which, as an edge-on galaxy, is likely to have a rather large fraction
of its $K$ band flux extincted relative to more face-on galaxies.  The
integrated 1.4 GHz flux from M~82 is about 8.4 Jy (White \& Becker
1992), while its $K$-band magnitude from 2MASS (Skrutskie et al. 2006)
is 4.7.  Scaling from these numbers, we can expect $K<12$ for nearly
any starburst galaxy we detect with NVSS.

\subsubsection{Clusters of galaxies}
Clusters of galaxies can also show strong emission at both X-rays and
radio.  A fraction of clusters of galaxies with strong radio haloes
shows $L_R$ of about $5\times10^{30}-5\times10^{32}$ ergs/sec/Hz, with
a strong relation between cluster radio and X-ray luminosity over the
range from $10^{44.5}-10^{45.5}$ ergs/sec, albeit with a rather small
sample (Cassano 2009).  The ratios of radio luminosity densities to
X-ray luminosities are thus about $10^{-13}-10^{-14}$ Hz$^{-1}$,
meaning that all the sources we have detected are too radio loud to be
clusters of galaxies.  The characteristic signature would be the
optical or infrared detection of a few of the galaxies in the cluster.
Additionally, one would expect that either the clusters are at
$z\gtsim0.5$, or that the radio emission would be resolved in NVSS,
since the clusters' radio haloes are typically $\sim 1$ Mpc in size
(Cassano 2009) -- and furthermore, the clusters would be expected
always to be resolved in the {\it Chandra} data.  One might also
expect to detect the brighter galaxies in a cluster in the infrared
data if a cluster of galaxies were the emitter.  E.g. M~87, the
central galaxy of the Virgo Cluster, has an absolute $K$-band
magnitude of -25.3 (Skrutsie et al. 2006), so even at redshift 0.5, it
would be at a magnitude of about $K=16.7$, with K-corrections applied
from Poggianti (1997).  The brightest few galaxies in a rich cluster
should thus be detectable in the VVV data.

\subsection{Unknown source classes and transients}
Some fraction of the sources in the GBS region may belong to
unidentified source classes.  For example, a few engimatic radio
transients have been seen from surveys taken near the Galactic Center
(e.g. Hyman et al. 2005; Hyman et al. 2009).  These transients have
been detected only at relatively low radio frequencies (235 MHz at the
Giant Metre-Wave Radio Telescope and 330 MHz at the VLA), but upper
limits at 610 MHz for the more recently detected of these transients
indicate that it has a spectrum steeper than $\nu^{-2}$ (Hyman et
al. 2009).  Since this object peaked at about 60 mJy at 235 MHz, it
would have been at no more than about 2 mJy at 1.4 GHz, meaning that
it would have been, at best, marginally detected in the NVSS data.
Such transients should thus be rather uncommon in the NVSS data set --
and a search for sources in only one of NVSS and FIRST (White et
al. 1997) showed that radio transients are likely to be quite rare at
1.4 GHz in single epoch searches (Levinson et al. 2002).

At the same time, the radio and X-ray data are taken far apart in time
(with the NVSS data taken mostly in 1993-6, and the GBS data taken in
2008 and 2009).  Therefore, variability can add some scatter to the ratios of
radio to X-ray luminosities.  In principle, such variability could
lead to misidentifications for a small fraction of the sources if they
are highly variable, but the source classes discussed here are either
low duty cycle in variability, or have small amplitudes of
variability, so this is a rather unlikely possibility.

\section{The matching sources and their identifications}
For the 12 sources which are good candidates for having radio
counterparts from NVSS, we present a case-by-case discussion of the
facts at hand revelant to the source's classification.  In particular,
we discuss other radio surveys, and we also discuss correlations with
public infrared surveys and with our own optical variability studies
and spectroscopic measurements as described above.  We draw attention
to the positional offsets of radio sources only for the cases where
the offsets are large enough that the matches may be spurious.

\subsection{CX~2 = CXO~J173728.3-290802, a known Seyfert 1 galaxy}
CX~2 is a known Seyfert 1 galaxy at a redshift of 0.0214 (Marti et
al. 1998).  Its radio flux density is $47\pm1$ mJy and its X-ray
measurements are affected by pile-up.  Its VVV infrared detections are
$J=13.22\pm0.01, H=11.90\pm0.01,$ and $K_s=10.94\pm0.01$.  The 2MASS
magnitudes for this source are $J=13.63\pm0.09, H=12.26\pm0.09$, and
$K_s=11.19\pm0.05$, and are likely to be more reliable than the
pipeline VVV magnitudes for sources this bright.  This object also
shows optical variability of about 0.2 magnitudes in $r'$ in an 7
night long campaign (see figure 2).  The absolute flux-scale
calibration for the r' variability survey has not yet been completed,
so the actual magnitude of the source in $r'$ remains uncertain by the
$\sim0.5$ magnitudes uncertainty that can be obtained by calibrating
against the USNO catalog; the best estimate for the mean value is 18.4.

\begin{figure}
\includegraphics[width=6 cm, angle=90]{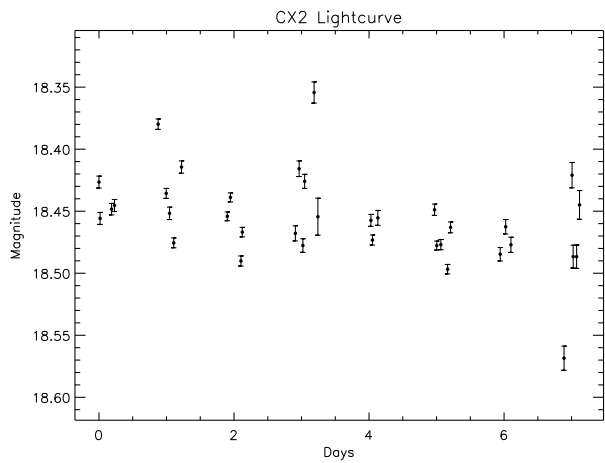}
\caption{The $r'$ light curve for CX~2.  The error bars represent the uncertainties in the relatively photometry, showing that the source clearly has irregular variability of at least about 0.2 magnitudes.  The absolute photometry is still calibrated only against USNO.  The variability is consistent with the AGN identification from existing optical spectroscopy.}
\end{figure}

\subsection{CX~40=CXO~J174404.3-260925, a new spectroscopically verified AGN}
This source is detected with 35 X-ray counts and a radio flux density
of $7.3\pm$0.5 mJy in NVSS.  The ratio of radio flux density to X-ray
flux is thus about $10^{-12.5}$ Hz$^{-1}$.  The source is also
detected in the Parkes-MIT-NRAO (PMN) 4.85 GHz radio survey's tropical
region (Griffith et al. 1994) with a flux density of $62\pm11$ mJy,
and a positional offset of about 62'', within the positional
uncertainties at this signal-to-noise and at this frequency from
Parkes.  Nonetheless, both the detection from the PMN survey and its
match to the GBS/NVSS source are marginal, so the 4.85 GHz radio
source may be spurious or unrelated.  If the match is taken to be real
then the radio source must be strongly variable, or strongly
self-absorbed, or a substantial fraction of the 1.4 GHz flux must have
been resolved out by the NVSS, as the spectral index implied by the
ratio of flux densities at the two frequencies is about +2.  The
source is detected in the VVV data with a positional offset less than
0.1'', at $J=15.42\pm0.03$, $H=14.21\pm0.03$ and $K_s=13.56\pm0.02$.
Our optical variability survey shows no evidence for variability from
this position, with the source at an $r'$ magnitude of $22.9\pm0.5$.

We have obtained an optical spectrum of this object (see figure \ref{cx40spec}).  Since this object is located
very close, on the sky, to a field star, a careful extraction was done
in IRAF to maximize the contribution from the AGN relative to that of
the star.  The optical spectrum for this source shows a series of strong lines and blends of strong lines.  The redshift for this object is determined primarily from the [O I] line at 6300 \AA (rest frame),
the only unblended line in the spectrum.  The
uncertainty of this redshift is difficult to estimate, since the only
unblended line we detect at high signal to noise is the [O I] line
which peaks at 6577 \AA in the observed frame -- rather close to the
wavelength of $H\alpha$ in the rest frame, which is potential
problematic given the nearby foreground star.  Regardless, the
uncertainty in the redshift should not affect the conclusion that the
source is an active galactic nucleus.

The full width at zero intensity (FWZI) of the H$\alpha$+[N II]
blend is 4400 km/sec, while the full width half-maximum (FWHM) for the
same blend is about 1470 km/sec.  The presence of telluric lines from
the telluric ``B'' band in the same range of wavelengths as the
H$\alpha$+[N II] blend is a cause for caution in taking too seriously
any conclusions which depend strongly on the profile of this line, and
especially its FWZI.  The isolated [O I] line has a FWHM of 910
km/sec.  It is likely that the FWHM of H$\alpha$ itself is
signficantly less than the FWHM of the blend, but the extent to which
this is true cannot be quantified with these spectra.  Regardless, the
data suggest a Seyfert 2 interpretation of the source, but because of
the foreground reddening to the source, the ratio of line strengths
of [O III] to H$\beta$ cannot be used as an additional discriminator
of source type.

The X-ray luminosity of the source, without accounting for foreground
absorption, would then be $4\times10^{42}$ ergs/sec in the 0.3-8 keV
band for $H_0=72$ km/sec/Mpc, $\Omega_{matter}=0.27$ and
$\Omega_{\lambda}=0.73$ (Spergel et al. 2003).  This luminosity is at
the lower end of the range typical of Seyfert galaxies (e.g. Kraemer
et al. 2004), but we note that the absorption is likely to be quite
strong, and has not yet been taken into account; with only 35 X-ray
counts it is difficult to estimate the appropriate correction from the
data.

\begin{figure}
\includegraphics[width=6 cm]{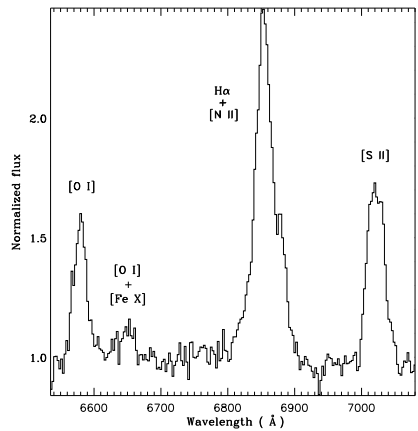}
\caption{The optical spectrum for CX~40.  The lines and their rest-frame wavelengths, from shortest to longest wavelengths, are [O I] 6300 \AA; a blend of [OI] 6364 \AA + [Fe X] 6375 \AA; a blend of H$\alpha$ 6562.76 \AA + [NII] 6548,6583 \AA, and a blend of [SII] 6716,6731 \AA.}
\label{cx40spec}
\end{figure}

\subsection{CX~49 = CXO~J173146.8-300309, a candidate AGN}
This source is detected at a flux density of 90$\pm$2.8 mJy in the
NVSS data and with 30 counts by Chandra.  Its ratio of radio flux
density to X-ray flux is thus about $10^{-11.4}$ Hz$^{-1}$.  The
source is detected in the VVV data with an offset of 0.21'', at
$H=14.22\pm0.03$ and $K_s=13.45\pm0.02$.  $J$-band data are missing in
the region around this source position.  This object also shows about
0.5 magnitudes of variability in the r' variability survey (see figure
4), with a typical value of $20.6\pm0.5$.

\begin{figure}
\includegraphics[width=5 cm, angle=90]{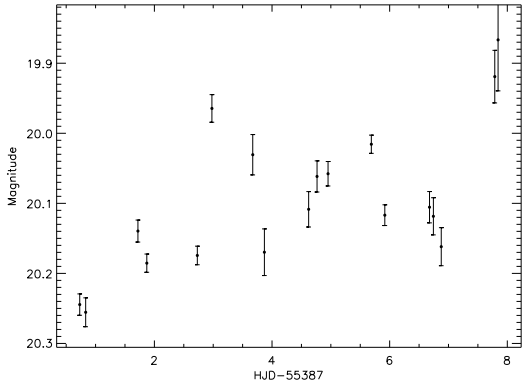}
\caption{The $r'$ light curve for CX~49.  The error bars represent the uncertainties in the relatively photometry, showing that the source clearly has irregular variability of about 0.5 magnitudes.  The absolute photometry is still calibrated only against USNO.  The variability is consistent with the AGN identification from its likely extended radio emission.}
\end{figure}

A source near this position is also detected in the zenith region of
the PMN survey (Wright et al. 1996), with a 4.85 GHz flux density of
123$\pm$14 mJy, and a positional offset of 98'', and is flagged as a
likely extended source (although there is no evidence for extension
from the NVSS catalog).  Given that there is no NVSS source nearer to
the PMN source than the one we have associated with CX~49, we regard
the match as likely reliable.  If the source is extended on the
spatial scales of the 4.2' beam size of the Parkes data, then likely
the VLA resolved out some of the flux from the source in the NVSS data
set, and a spectral index cannot be estimated from these data, but
otherwise, the source has a flat to slightly inverted spectrum.  The
possible extended radio emission and bright infrared counterpart both
suggest that the source is an active galactic nucleus, but the
identification is insecure at this point.

\subsection{CX~52 = CXO~J174423.5-311636, a likely AGN}
This source is detected at a flux density of $425\pm15$ mJy in NVSS,
and with 29 counts by Chandra, giving a ratio of radio flux density to
X-ray flux of about $10^{-10.7}$ Hz$^{-1}$.  There are no optical or
infrared stars near this source, indicating that it is likely in a
region of high extinction -- the nearest sources in the VVV catalog
are more than 3'' away, and all have very red colors; not
surprisingly, the source is not detected in our optical variability
survey.  The source's X-ray hardness ratio is 0.83$\pm0.26$, making it
likely one of the hardest sources in the GBS, and while, with the
relatively small number of counts, this designation cannot be secure,
this finding supports the idea that the source is in a region of large
foreground extinction.  It is also detected at a flux density of 882
MHz in the Galactic Center 330 MHz survey of Nord et al. (2004), but
with the VLA in A\&B configuration, so that the beam size of those
data is considerably smaller than for the NVSS data.  Roy et
al. (2005), using imaging data from the VLA and the Australia
Telescope Compact Array with higher angular resolution than the NVSS
data identify this source as a compact core plus extended emission,
and hence an extragalactic object.  A likely match is also found in
the Nobeyama 10 GHz survey (Handa et al. 1987) -- source number 4 in
that catalog is about 2' from CX~52, and appears pointlike in that
survey with a flux density of 0.46$\pm0.2$ Jy.  The stated positional
accuracy of the Nobeyama data is about 5'', but the beam size for that
survey was about 3', and the offset may be the result of systematic
errors in the source centroiding, especially given that the source
appears to be extended in lower frequency data, and could perhaps be
mildly extended and asymmetric in the 10 GHz data, perhaps leading to
an offset against the X-ray position.  The lack of any other sources
in the first catalog within 3' of the Nobeyama position strongly
supports this possibility.

\subsection{CX~233 = CXO~J174206.1-264117}
This source is detected at a flux density of 3.5$\pm0.5$ mJy in NVSS,
and with 10 X-ray counts, yielding $L_R/L_X$ of $10^{-12.3}$
Hz$^{-1}$.  The match is that with the largest separation of any match
presented here, 5.0'', but given the faintness of the NVSS source and
its suggestive evidence for being extended (the major axis in the NVSS
catalog is inferred to be about 78''), the NVSS positional uncertainty
is about 10'', so the positions are certainly consistent with one
another. The nearest infrared source in VVV is 1.6'' away, and has
$J=17.48\pm0.24$, $H=16.27\pm0.17$ and $K_s=15.68\pm0.15$.  Given that
the source is far off-axis, the 95\% positional uncertainty estimate
based on the formula of Hong et al. (2005), developed to deal with the
point-spread function of Chandra, is 6.1'' The position of the
potential infrared match is thus consistent with the X-ray source
position.  The spatial extent of the radio counterpart is suggestive
evidence that the source is an active galactic nucleus.  The optical
variability survey shows a source wirh $r'=23.0\pm0.5$, but no
evidence for variability.

\subsection{CX~293 = CXO~J174000.6-274816, a likely AGN}
This source is detected at a flux density of 170$\pm6$ mJy in NVSS,
and with 9 X-ray counts, yielding $L_R/L_X$ of $10^{-10.7}$ Hz$^{-1}$.
It is also detected in the 330 MHz survey of Nord et al. (2004), with
a flux density of 404 mJy.  The source is identified in the radio
imaging of Roy et al (2004) as a triple source, and hence is likely to
be an active galactic nucleus.  It has a VVV infrared counterpart
1.0'' away, with $J=14.80\pm0.02$, $H=13.75\pm0.02$ and
$K_s=13.33\pm0.03$.  An optical spectrum was taken of this VVV source
and showed a purely stellar spectrum with no emission lines,
indicating that it is likely a chance superposition (see figure
\ref{cx293}).  A 10\% flux drop is seen around 6564 \AA, so it is
clear that the source is a Galactic foreground object rather than,
e.g. a background blazar, but most of the strong features in the
spectrum correspond to atmospheric features that have not been removed
fully.  Additionally, the astrometry from our optical variability
survey indicates that the source of which the spectrum was taken was
probably not the real match.

\begin{figure}
\includegraphics[width=6 cm]{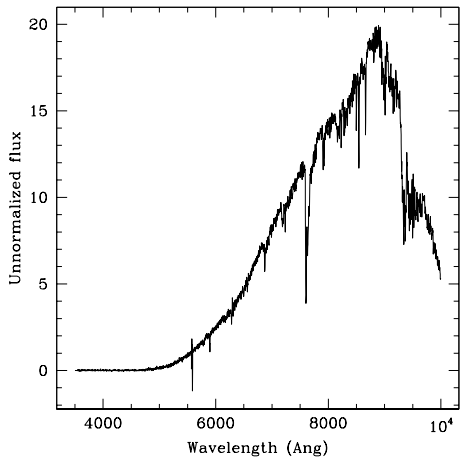}
\caption{The optical spectrum for CX~293.  The strong features
are due to imperfect removal of atmospheric features.}
\label{cx40spec}
\end{figure}

\subsection{CX~390 = CXO~J173607.5-294858}
The source is detected at a flux density of 29$\pm1$ mJy in NVSS and
with 7 X-ray counts giving $L_R/L_X$ of $10^{-11.3}$ Hz$^{-1}$.  The
source could be a spurious match, as the offset between the X-ray and
radio positions is 3.1'', despite the radio position being accurate to
about 1.4'', although the 95\% error circle on the X-ray position
extends out to about 3.2'', using the formula of Hong et al. (2005).
The X-ray source has a match in the VVV data 0.71'' away, with
$J=17.56\pm0.25$, $H=16.50\pm0.25$ and $K_s=15.92\pm0.23$.  The
optical counterpart of this object has $r'=22.6\pm0.5$ and shows no
evidence for variability.

\subsection{CX~488 = CXO~J173605.3-283232}
The source is detected at a flux density of 126$\pm5$ mJy in the NVSS
data and with 6 X-ray counts, giving $L_R/L_X$ of $10^{-10.6}$
Hz$^{-1}$.  The source is found at a flux density of 707 mJy in the
330 MHz survey of Nord et al. (2004).  The spectrum of this source is
steeper than $\alpha=-1$, which is the spectral criterion Nord et
al. (2004) used for identifying new pulsar candidates.  This source
was excluded as a pulsar candidate on the basis of its having a major
axis of 15.9'', just larger than the threshold of 15'' used by Nord et
al. (2004) to identify extended sources.  The X-ray source position
has a match in VVV at $H$ and $K_s$, with magnitudes 15.44$\pm0.09$
and $14.94\pm0.08$, respectively, with a positional offset of 1.2'',
which casts some doubt on the possibility that the source is a pulsar,
but if the infrared counterpart is a chance superposition, the source
is a good pulsar candidate.  The optical counterpart in our
variability survey has $r'=21.4\pm0.5$, and shows no evidence for
variability.

\subsection{CX~494 = CXO~J173458.8-301328}
The source is detected at a flux density of 29$\pm1$ mJy in the NVSS
data and with 6 X-ray counts, giving $L_R/L_X$ of $10^{-11.3}$
Hz$^{-1}$.  A near infrared counterpart is found 1.8'' away (but given
the 0.8'' statistical uncertainty on the GBS position, this is
reasonably likely to be a real match), and has $J=13.74\pm0.01$,
$H=12.90\pm0.01$ and $K_s=12.51\pm0.01$.  Its 2MASS magnitudes are
$J=13.62\pm0.07, H=12.71\pm0.08$, and $K_s=12.35\pm0.04$, and may be
more reliable than those from VVV because of the brightness of the
object -- although we note that the differences are small.  There is
no variability detected in the optical variability survey, but a
source is seen at the position.  There are two possible optical
counterparts in our variability survey, neither of which varies.  They
have $r'$ magnitudes of 17.3$\pm0.5$ and 18.3$\pm0.5$.

\subsection{CX~578 = CXO~J174442.3-311633}
The source is detected at a flux density of 2.8$\pm0.6$ mJy in the
NVSS data and with 5 X-ray counts, giving $L_R/L_X$ of $10^{-12.2}$
Hz$^{-1}$.  The offset between the X-ray and radio positions is about
3.1'', but with the nearly 20'' positional uncertainty on the radio
source (due both to its faintness, and its large major axis of 122''
according to the NVSS catalog), the positions are quite consistent
with one another, despite the relatively large offset.  The nearest
infrared counterpart is 1.7'' away, which makes it a marginal match,
given the good (0.3'' statistical uncertainty, 0.6'' boresight
uncertainty) X-ray position; the 95\% uncertainty on the X-ray source
position from Hong et al. (2005) is 1.2''.  It has $J=16.74\pm0.04$,
$H=14.86\pm0.03$ and $K_s=14.19\pm0.03$.  In our optical survey, the
source is detected as a $r'=20.7\pm0.5$ star, with no evidence for
variability.

\subsection{CX~937 = CXO~J175359.2-281720, a likely AGN}
This source is detected with a flux density of $52\pm2$ mJy and with 3
X-ray counts, yielding a $L_R/L_X$ of $10^{-10.1}$ Hz$^{-1}$.  Roy et
al. (2005) identify this source as a background active galactic
nucleus because it is resolved into a double source with the high
resolution radio imaging data.  The source is identified in the PMN
survey (Wright et al. 1996) as a 104$\pm$12 mJy source at 4.85 GHz.
Given that Roy et al. (2005) identify the source as extended, it is
most likely that some radio flux is resolved out in the NVSS data, and
that the spectrum is not really inverted.  It is also identified at
186 mJy in the 330 MHz data set of Nord et al. (2004), with the
spectral index implied from comparing the 330 MHz data with the NVSS
data consistent with the expectations for a typical background active
galactic nucleus.  A bright infrared counterpart is seen 1.1'' away,
with $J=13.68\pm0.01$, $H=13.39\pm0.02$ and $K_s=13.32\pm0.03$ from
VVV, and with $J=13.82\pm0.1$ from 2MASS, with the 2MASS photometry at
$H$ and $K_s$ flagged as unreliable.  The source is saturated in our
optical variability survey, making searches for variability from it
problematic.

\subsection{CX~1234 = CXO~J173531.4-295145}
This source is detected with a flux density of $4.7\pm0.5$ mJy and
with 3 X-ray counts, yielding a $L_R/L_X$ of $10^{-11.3}$ Hz$^{-1}$.
The match offset is about 3.3'', well within the positional accuracy
of the NVSS source, but at a larger separation than most of the
matches in this paper.  The nearest infrared counterpart is located
1.4'' from the X-ray position, and is undetected in $J$, but found at
$H=16.19\pm0.2$ and $K_s=15.84\pm0.22$.  In our optical variability
survey, the source is swamped out by a bright, saturated star 5''
away.

\begin{table*}
\begin{tabular}{lrrrrl}
\hline
GBS cat. no& $F_X$ & $S_{1.4}$& $K_s$ & log($L_R/L_X$)& Comments/classification\\
\hline
2& ** &47 & 10.94& **& known AGN, X-rays affected by pileup\\
40& $2.8\times10^{-13}$& 7.3& 13.56 & -12.5 & spec. ver. AGN\\
49& $2.4\times10^{-13}$& 90& 13.45& -11.4 & opt. var.\\
52& $2.3\times10^{-13}$&425& non-det& -10.7& AGN from ext. emission\\
233& $8\times10^{-14}$&3.5&15.68& -12.3& large offset, but large pos. uncer.\\
293&$7\times10^{-14}$&170&13.33&-10.7&triple source, likely AGN\\
390&$6\times10^{-14}$&29&15.92&-11.3&least likely match\\
488&$5\times10^{-14}$&126&14.94&-10.6&marginal extension at 330 MHz may rule out pulsar\\
494&$5\times10^{-14}$&29&12.51?&-11&\\
578&$4\times10^{-14}$&2.8&14.19?&-12.2&poss. ext. in NVSS\\
937&$2\times10^{-14}$&52&13.32&-10.1&double radio source, prob. AGN\\
1234&$2\times10^{-14}$&4.7&15.84&-11.3&\\

\end{tabular}
\caption{Key parameters and attempted classifications for the NVSS
sources in the GBS region.  The columns are (1) GBS catalog number (2)
X-ray flux in ergs/sec/cm$^2$ (3) radio flux density from NVSS in mJy
(4) $K$ band magnitude from VVV (5) ratio of radio luminosity density
to X-ray luminosity in Hz$^{-1}$ and (6) any key additional
information about the sources.  Infrared matches with question marks
are at separations more than 1.5''.  In many of these cases, the
matches may still be good, since these large offset matches are all
for the X-ray sources with fewer than 10 counts, and hence relatively
poorly constrained positions.}
\end{table*}

\section{Discussion}
We have considered the multi-wavelength properties of 12 X-ray sources
in the Chandra Galactic Bulge survey with radio counterparts in NVSS.
The lack of any very bright infrared counterparts to the X-ray/radio
matches argues against any of these objects' being pre-main sequence
stars, massive stars (either single or in binaries) or symbiotic
stars.  Additionally, massive stars would likely be Tycho sources
unless they were highly absorbed, and none of these objects matches
with the Tycho catalog (R.I. Hynes et al. in prep.).  The sources are
all overluminous in the radio for being coronally active stars, unless
they are ultracool dwarfs, or for being X-ray binaries or cataclysmic
variables or background clusters of galaxies.  Coronally active
ultracool dwarfs can be ruled out on the basis of the lack of bright
infrared counterparts, unless their proper motions are so large that
the stars have left the matching regions we have used with respect to
2MASS.  Starburst galaxies are also unlikely, unless the galaxies have
even larger ratios of radio flux to infrared flux than that of M~82.

The lack of sources in the aforementioned classes in the current list
of radio/X-ray matches does not preclude their existence among the GBS
sources -- some of the sources not presently detected in the radio are
likely to be coronally active binaries, for example.  Some clear
examples of X-ray binaries are already known in the survey (J11).  On
the other hand, pulsar wind nebulae and supernova remnants can likely
be ruled out throughout the survey region as they would be bright
X-ray sources which would have been detected in the NVSS data as well,
and this is relatively unsurprising given that the field is outside
the Plane of the galaxy, so any such objects would have to be either
very nearby or among the rare young objects at large scale-height.

The only viable classifications for the sources detected, then, are
active galactic nuclei and pulsars, plus planetary nebulae -- the two
classes which span very large ranges in all flux ratios, and the
planetary nebulae which have systematically higher ratios of radio to
X-ray fluxes than the other source classes.  Planetary nebula origins
are disfavoured as well.  Nearly all the sources have either too low a
ratio of $L_R/L_X$, or have radio morphology or optical variability
that argue against PN origins.  CX~488 has an especially steep radio
spectrum not consistent with the free-free emission process that
dominates the radio emission from PNe.  CX~494 is a viable candidate,
which could be similar to the X-ray brightest PN, BD+30$^\circ$3639.
If this can be verified, then the source should be strongly
emission-line dominated in optical and infrared spectra.  At the
present time, we do not have an optical nor infrared spectrum of this
object.  CX~1234 has a ratio of X-ray to radio flux which is
reasonable, but its infrared flux is too faint compared to the X-ray
flux to be a PNe unless it is an extreme outlier.

Two of the objects have been definitely classified as active galactic
nuclei by spectroscopic follow-up (CX~2 -- Marti et al. 1998 and CX~40
-- this paper).  Several others (CX~52, CX~293, and CX~937) show
morphology in high angular resolution radio data that indicates that
they are likely background active galactic nuclei (Roy et al. 2005),
and the optical variability of CX~49 argues for an AGN nature for the
source, given that all the classes of stellar objects which would show
optical variability are ruled out by the large ratio of radio to X-ray
fluxes.  CX~488 represents a viable candidate for being a pulsar,
although the marginal extension reported in Nord et al. (2005) does
cast doubt on this interpretation.

Some of these background AGN can serve as important probes of the
interstellar medium of our Galaxy.  Many of them are bright enough for
high resolution spectroscopy in the near infrared and/or optical,
meaning that searches for absorption lines in their spectra can be
used to estimate the metal abundance of the interstellar medium of the
Galaxy in the inner part of the Bulge.  Additionally, it has already
been noted (Roy et al. 2005) that AGN behind the inner Galaxy can be
used to investigate the strength and structure of the inner Galaxy's
magnetic field through radio Faraday rotation measurements.

Additionally, this method has proved to be extremely effective in
identifying the background AGN in the GBS survey.  The estimated
number of AGN expected in the full survey was about 25 (J11), so with
3/4 of the survey taken, about 18 AGN would be expected.  In
principle, if the number of AGN is below the expected number of AGN
(e.g. due to Poisson variations) and our still unclassified radio
sources do turn out to be active galactic nuclei, we could have
detected all the AGN in the survey already, but this is unlikely given
that AGN can be $\sim1000$ times too faint in the radio to have been
detected in these data if they are at the faint end of the X-ray flux
distribution and at the radio faint end of the radio to X-ray flux
ratio distribution, but it is likely that from the radio correlations we have already identified more than half of the AGN in the survey.  

The addition of the NVSS data to the GBS project has thus been of
significant value, even if it has not resulted in the identification
of the stellar objects that are the primary motivation for the survey.
It is also clear from the work here that in order to find substantial
numbers of stellar objects, a deeper radio survey of this region would
be needed.  Such a survey could be done easily now with the improved
sensitivity of the EVLA.

\section{Acknowledgments}

TJM thanks Anna Scaife for pointing out the existence of the high
frequency published radio surveys which overlap with the GBS field,
for making a visual inspection of the Nobeyama data for CX~52 and for
discussion of the radio properties of protostars.  This publication
makes use of data products from the Two Micron All Sky Survey, which
is a joint project of the University of Massachusetts and the Infrared
Processing and Analysis Center/California Insitute of Technology,
funded by the National Aeronautics and Space Administration and the
National Science Foundation.  TJM and AD thank the Science and
Technology Facilities Council for support under a rolling grant to the
University of Southampton.  PGJ and GN acknowledge support from the
Netherlands Organization for Scientific Research.  RIH and CB
acknowledge supprt from National Science Foundation Grant AST-0908789.
DS acknowledges support from STFC through an Advanced Fellowship.  SG
acknowledges support through a Warwick Postgraduate Research
Scholarship.  RW is partially supported by a European Research Council
Starting Grant.

\label{lastpage}

\end{document}